\documentclass[useAMS,usenatbib]{mn2e}
\usepackage{graphicx}     
\usepackage{epsfig}  
\usepackage{times}    
\usepackage{subfigure}

\title[Steepened density profiles via group interactions]
  {Steepened inner density profiles of group galaxies via interactions: \\ An N-body analysis}
\author[B. M. Dobke et al.]
  {Benjamin M. Dobke,
  Lindsay  J. King, Michael Fellhauer\\
 Institute of Astronomy, University of Cambridge, Madingley Rd, Cambridge CB3 0HA\thanks{E-mail: bdobke@; ljk@; madf@; ast.cam.ac.uk}}
\date{Accepted ... / Received ... }

\pagerange{\pageref{firstpage}--\pageref{lastpage}} \pubyear{2007}

\begin{document}

\label{firstpage}

\maketitle

\begin{abstract}
We continue to see a range of values for the Hubble constant obtained from gravitationally lensed multiple image time delays when assuming an isothermal lens despite a robust value from the HST key project (72 $\pm$ 8 km\,s$^{-1}$ Mpc$^{-1}$).  One explanation is that there is a variation in Hubble constant values due to a fundamental heterogeneity in lens galaxies present in groups, i.e. central galaxies with a high dark matter surface density, and satellite galaxies with a possibly stripped halo, low dark matter surface density, and a more central concentrated matter distribution.  Our goal is to see if a variety of group interactions between the most massive group members can result in significant changes in the galaxy density profiles over the scale probed by strong lensing ($\la$15 kpc).  While stripping of the outer parts of the halo can be expected, the impact on inner regions where the luminous component is important is less clear in the context of lensing, though still crucial, as a steepened density profile within this inner region allows these lens systems to be consistent with current HST/WMAP estimates on $H_{0}$.  We employ the particle-mesh code SUPERBOX to carry out the group interaction simulations.  An important advantage of using such a code is that it implements a fast, low-storage FFT-algorithm allowing simulations with millions of particles on desk-top machines.  We simulate interactions between group members, comparing the density profile for the satellite before and after interaction for the mass range of $10^{11}$ to $10^{13}$ $\rm M_{\sun}$. Our investigations show a significant steepening of the density profile in the region of $\sim$ 5-20 kpc, i.e. that which dominates strong lensing in lens galaxies.  This effect is independent of the initial mass-to-light ratio.  Additionally, the steepening in the inner region is transient in nature, with consecutive interactions returning the profile to an isothermal state within a timeframe of $\sim$ 0.5 - 2.0 Gyr.  This factor may help explain why lens galaxies that produce lower values of $H_{0}$ (i.e. those with possibly steeper profiles) are far fewer in number than those which agree with both the HST key project value for $H_{0}$ and isothermality, since one would have to observe the lens galaxy during this transient steepened phase.
\end{abstract}


\section{Introduction}
A range of values for the Hubble constant is obtained from different strong gravitational lens systems with measured image time delays, assuming isothermal models for the lensing galaxies (e.g. \citealt{bigg}; \citealt{gilm}). The estimates are typically lower than the values from the HST key project and WMAP which are consistent with each other at 72 $\pm$ 8 km\,s$^{-1}$ Mpc$^{-1}$ and 72 $\pm$ 5 km\,s$^{-1}$ Mpc$^{-1}$, respectively \citep{free,sper}.  Since this Hubble constant value is considered robust, this variation raises questions as to the relative extent of the galaxy halo and the luminous component since the derived values fundamentally depend on the assumed lens potential and matter surface densities.  A number of differently motivated studies have found that the inner regions of lens galaxies exhibit isothermal density profiles (e.g. \citealt{rusi}), including recent results from the Sloan SLACS survey \citep{koop} which found $\eta$ = 2.01 $^{+0.02}_{-0.03}$ (1$\sigma$; $\rho_{\rm tot}$ $\propto$ $\rm$ r$^{-\eta}$) when studying 15 early-type galaxies.  Moreover, studies inferring the Hubble constant from 10 time-delay lens galaxies produced values in agreement with the HST/WMAP value, while still being consistent with simulated time-delays from N-body galaxies which were isothermal in nature \citep{sahaa}.  Even a relatively simple analysis using the existing time delay systems indicates a consistency with isothermality \citep{dobke}.  Stellar dynamical studies also point towards isothermal density profiles in the inner regions of galaxies  (e.g.  \citealt{rix}; \citealt{roma}; \citealt{gerh}; \citealt{treua}).

Despite this apparent trend towards isothermality, we still observe a number of systems where it is hard to reconcile $H_{0}$ values with the HST key project value when assuming an isothermal model. Modelling of the systems PG1115+080, B1600+434, HE2149-2745, and SBS1520+530 has shown that these systems prefer more centrally concentrated density profiles and hence non-isothermal models \citep{kocha, kochc}.  In the specific case of PG1115+080, detailed modelling done by combining lensing, stellar kinematic and mass-to-light ratio constraints, in order to build a two-component model, shows a mass density profile significantly steeper than isothermal at $\eta$ = 2.35 $\pm$ 0.15, where $\rho_{\rm tot}$ $\propto$ $\rm$ r$^{-\eta}$.  However, the derived value for the Hubble constant came to $H_{0}$ = 59 $^{+12}_{-7}$ km\,s$^{-1}$ Mpc$^{-1}$, implying the need for an even steeper profile to agree with the HST key project value \citep{treub}.

\citet{keet} have shown that lens models of four-image systems \emph{overestimate} the Hubble parameter by up to 15\% when neglecting the effect of the contribution from other group members on the potential of the lens galaxy, with an even higher discrepancy for two-image systems.  Hence accounting for environment would further reduce the derived value of $H_{0}$ for a fixed primary lens model, requiring the slope of the primary lens to be further steepened to obtain consistency with the HST key project value.  Additionally, outside the group environment, line-of sight structures can provide contributions to the lens potential \citep{momc}. 
 
While isothermal systems certainly appear to be the more prevalent, an explanation of this sub-sample of lens galaxies requiring non-isothermal profiles is not clear.  One suggestion focuses on the possible role that the group may have on the lens galaxy, specifically the effect of tidal stripping of satellite galaxies \citep{kochb} as recently observed in clusters using galaxy-galaxy lensing \citep{limo}.  Although a full census of lens galaxy environments is not available, a number of observations \citep{tonry, fass, will} seem to show that typical early-type lens galaxies should be members of a group.  In the halo model described by \citet{cooray}, one galaxy typically lies at the centre of the group halo, while other galaxies are smaller satellites orbiting in that halo.  In this scenario we have two differing cases for galaxy structure.  In the first we see the central galaxies with a high dark matter surface density and the luminous component making only a small fraction of the overall matter content.  In the second case, we see a satellite galaxy with a possibly stripped halo, and more centrally condensed core.  In essence, it suggests a fundamental heterogeneity in early-type galaxies present in groups, and when applied to lens galaxies the distinction between the two surface mass densities could give rise to a difference in the derived Hubble constant were a standard isothermal model assumed.  

This description of the heterogeneity in groups begs a simple question;  how much can the group members actually effect each other in terms of changing their mass and density profiles?  In such groups, unlike large clusters, there is not so much a single dominant central galaxy, but rather a power sharing scenario between two or more approximately equivalent galaxies.  While interactions between members with large differences in their respective masses should be expected to produce large dynamical and structural effects in the less massive member (e.g. ultracompact dwarf (UCD) interactions; see \citealt{fella}), the effect of interactions between more similarly matched group members is perhaps not so clear when considering the effect in the context of lens galaxies.  Indeed, although certain past studies (e.g. \citealt{haya}; \citealt{kaza}) have investigated the effect of satellite-halo interactions, the focus has been on the evolution of substructure or of the very inner cusp rather than the effect on cosmological parameter derivation from lens galaxies.

The question posed is a complex one, and the parameter space of group interactions is vast.  In this paper our aim is to establish whether, over the scale probed by strong lensing ($\la$15 kpc), significant changes in the density profile of a galaxy on a satellite orbit can be induced by interactions with a group galaxy at a central locale.  While stripping of the outer parts of the satellite's halo could be easily expected, the impact on inner regions where the luminous component is important is less clear in the context of lensing, though crucial, as it is with a steepened density profile within this inner region that the time delays for these lens systems can agree with current HST/WMAP estimates on $H_{0}$.  In this way we ascertain if another mechanism is required to alter density profiles, possibly during the group's formation.

The outline of the paper is as follows;  in \S2 we briefly introduce the N-body code used to simulate the interactions, with details of the exact parameters of the simulated objects.  In \S3, we go on to present results and discuss their analysis in the framework of the problem.  Finally, \S4 draws some conclusions based upon our findings.

\begin{figure*}
\centering
\epsfig{file=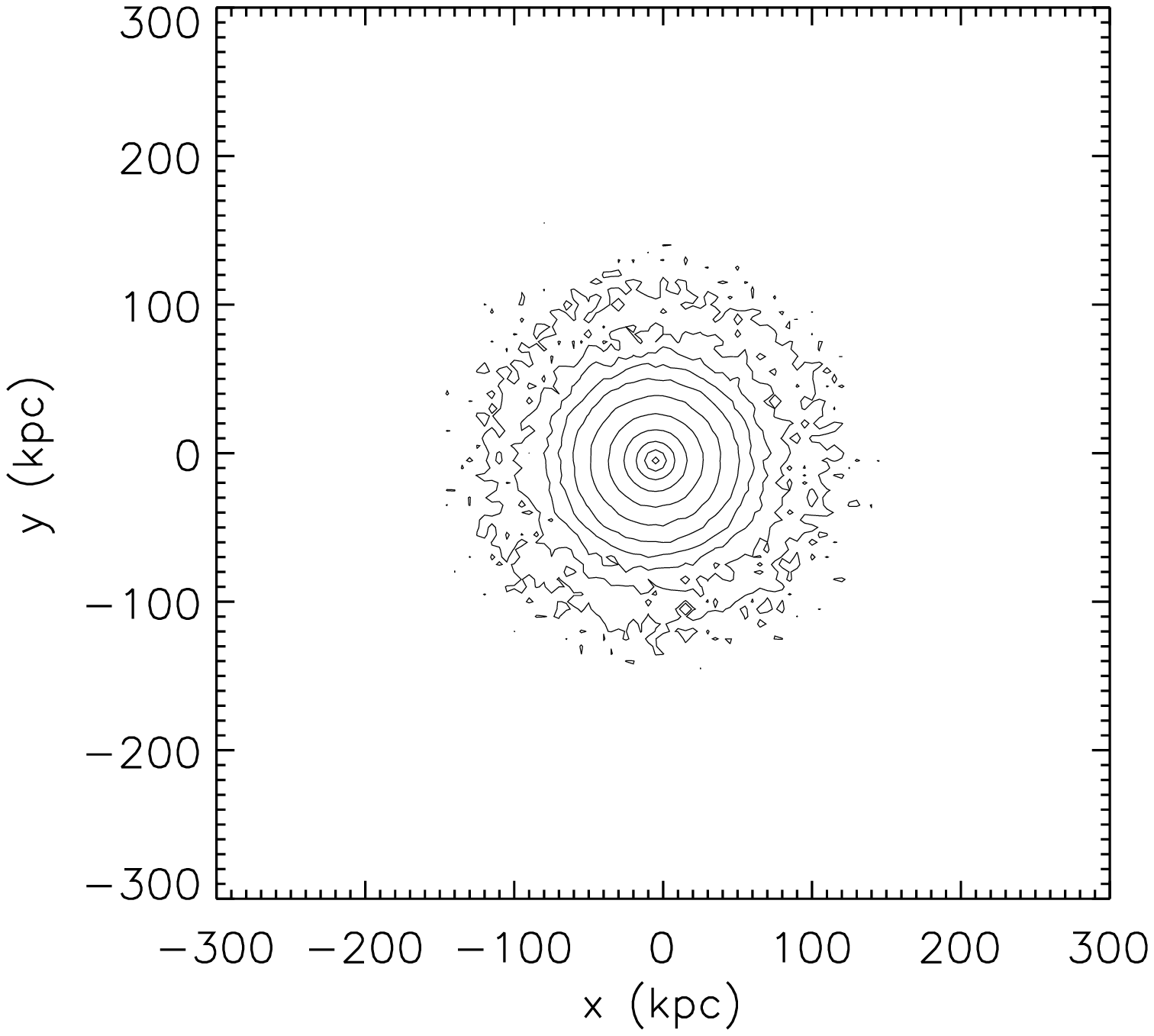,scale=0.43}
\epsfig{file=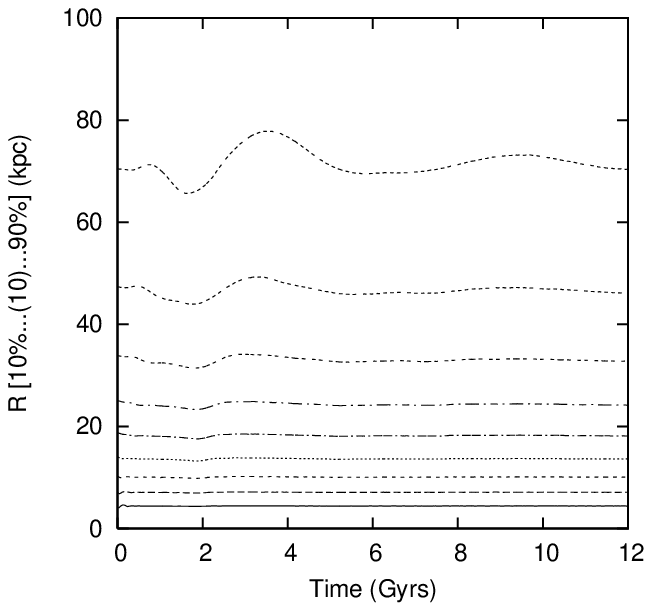,scale=1.00}
\caption{
{\sl Left:}  Contour plot of an NFW halo after undergoing isolated integration.  The contours have (logarithmic) magnitude spacing with the innermost and outermost contours corresponding to 16 and 36 mag $\mathrm{arcsec^{-2}}$ respectively.  The object shows clearly defined symmetrical structure and is ready to be used in a simulation scenario.  {\sl Right:}  The evolution to dynamic equilibrium as shown by the Lagrangian-radii of the mass shells of 10\% to 90\% of particles in a typical simulation.  The flattening of the lines with time shows that an equilibrium state has been reached, and that the object is ready to be used in an interaction scenario.}
\label{plot1}
\end{figure*}

\begin{figure*}
\centering
\epsfig{file=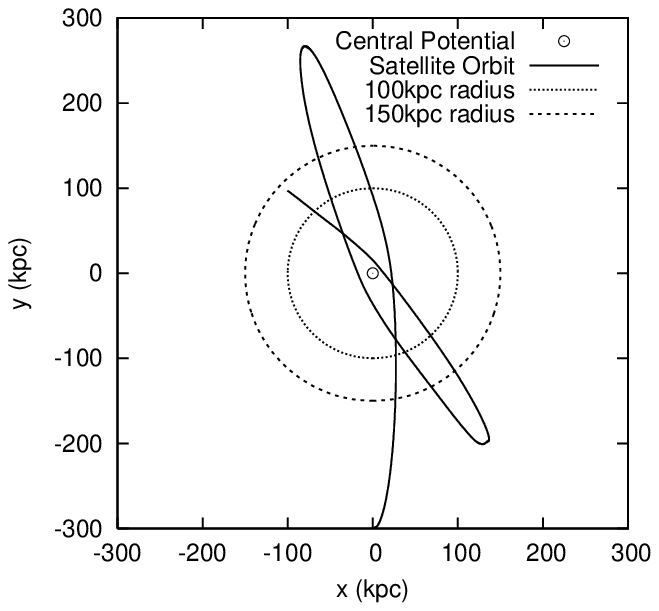,scale=1.00}
\epsfig{file=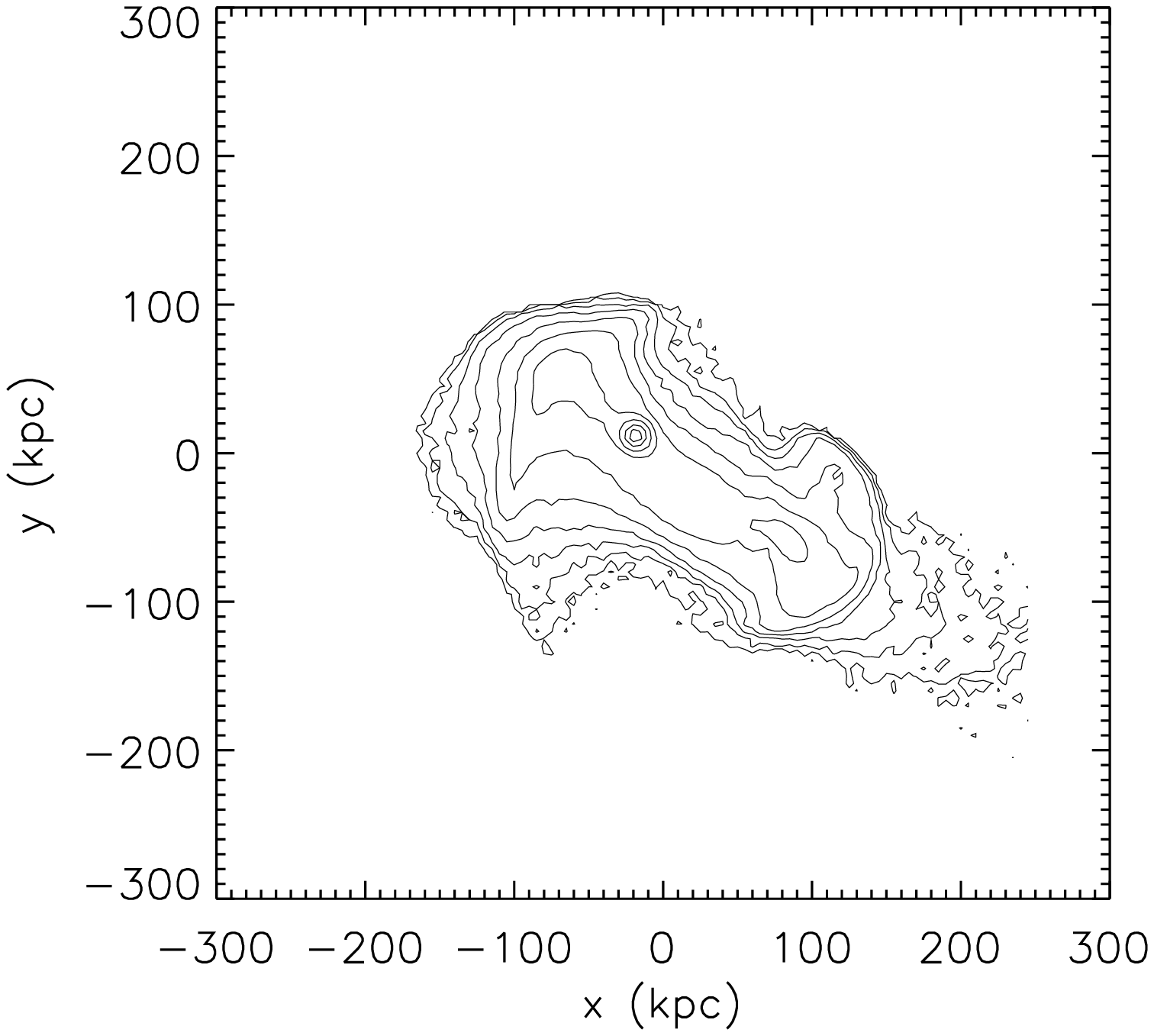,scale=0.43}
\caption{
{\sl Left:} Example of a path taken in the x-y plane during a simulation run.  The orbit starting position in the x-y plane is (0,-300) kpc, with a tangential velocity $V_{tan}$ = +100 km\,s$^{-1}$ in the x-dimension.  This particular run shows the interaction between a $10^{11}$ $\rm M_{\sun}$ object after undergoing three passes of a central $10^{12}$ $\rm M_{\sun}$ object.  The run time is 7.3 $\times$ $10^{9}$ years.  Also shown are the 100 kpc and 150 kpc radial distances from the central potential, beyond which the satellite spends 78\% and 64\% of its orbit time respectively.  {\sl Right:}  Typical contour plot for the same $10^{11}$ $\rm M_{\sun}$ halo after undergoing two interactions with a $10^{12}$ $\rm M_{\sun}$ central object.  Note the formation of tidal-tail structures in the outer regions.}
\label{plot2}
\end{figure*}

\section{Theory and Simulation Setup}

\subsection{The particle-mesh code SUPERBOX}

Our analysis employs the particle-mesh code SUPERBOX to carry out the simulations.  The code utilises high resolution sub-grids and a nearest grid point (NGP) force-calculation scheme based on the second derivatives of the input potential.  An important advantage of using such a code is the fact that it implements a fast, low-storage FFT-algorithm giving the possibility to work with millions of particles on desk-top machines or small clusters.  

For each object used in the simulation, there are 5 grids with 3 different resolutions.  For the purposes of our simulation setup, we define the size of the inner, middle and outer grid sizes.  The inner grid is the highest-resolution grid which resolves the centre of the galaxy.  The middle grid has an intermediate resolution to resolve the galaxy as a whole, while the outer grid is the size of the whole simulation area (i.e. `the local universe'), and has the lowest resolution.   The inner and middle grid remain focused on the object throughout the simulation.  Each grid has a predefined number of mesh-points.  In our simulations we use $2^{6}$ mesh-points.  Doubling the number to $2^{7}$  showed no significant effect on the overall results discussed in \S3, indicating a high enough level of resolution.  For a full description of the code and grid structure see \citet{fellb}.

\subsection{Simulation Setup}

We focus the main body of our investigations on approximately imitating the dynamics and parameters of lens systems with a known history of lens models with low dark matter surface densities, and hence possible steeper density profiles,  e.g. PG1115+080, B1600+434, HE2149-2745, SBS1520+530 \citep{kocha}, and as such are clear candidates for possible satellite stripping.  We use the term satellite here to refer to a galaxy undergoing orbital interactions with another galaxy, whether that be of equivalent or higher mass, and not in the sense of e.g. a dwarf elliptical satellite.  Some of the parameters used to set up the model galaxies were taken in part from \citet{impey} and \citet{keet}.  Parameters such as mass and scale are obviously often model dependent and as such were only taken as approximate guide values and sanity checks.  In actuality, variations of up to an order of magnitude in the masses of the model objects had little effect on the overall results; more important were the relative masses between the interacting bodies.

\begin{table*}
\centering
\begin{minipage}{18.0cm}
\caption{Table of our initial model parameters for three particular interactions.  The first two rows correspond to the object parameters that were used in the simulations of Fig.\,\ref{plot3} and Fig.\,\ref{plot4}. We include NFW and Hernquist mass with corresponding scale and cut-off radii, the characteristic crossing time ($T_{cr}$) and the grid resolutions used in each case.  We use the virial radius as an approximate guide for the cut-off radii.  The middle grid size is then chosen to encompass this value.  Parameters for the $10^{13}$ $\rm M_{\sun}$ object were taken in part from \citet{mcla} and \citet{vesp}}
\label{table1}
\begin{tabular}[t!]{cccccccc}
\hline
\multicolumn{4}{|c|}{Initial Object Parameters}&\multicolumn{3}{|c|}{Corresponding Analytic Potential Parameters}&\multicolumn{1}{|c|}{Grid Resolution}\\
Mass$_{N,H}$ ($\rm M_{\sun}$)&$r_{s, N}$/$r_{s, H}$ (kpc)&$r_{c, N}$/$r_{c, H}$ (kpc)&$T_{cr}$ (Myr)&Mass$_{N,H}$ ($\rm M_{\sun}$)&$r_{s, N}$/$r_{s, H}$ (kpc)&$r_{c, N}$/$r_{c, H}$ (kpc)& In./Mid./Out. (kpc)\\
\hline
$10^{11}$/$10^{10}$&10/10&120/120&292.36&$10^{12}$/$10^{11}$&20/10&250/120&50/150/1000\\
$10^{12}$/$10^{11}$&20/10&250/120&250.58&$10^{12}$/$10^{11}$&20/10&250/120&50/300/1000\\
$10^{12}$/$10^{11}$&20/10&250/120&250.58&$10^{13}$/$10^{12}$&560/5&150/150& 50/300/1000\\

\hline
\end{tabular}
\end{minipage}
\end{table*} 

For the dark matter halo component of each of the group galaxies we use a Navarro, Frenk \& White (NFW) profile \citep{nav} while for the luminous component we use a Hernquist profile \citep{hern}.  Hence, the combined density profile for each of the modelled galaxies has the form,

\begin{eqnarray}
&&\nonumber\rho_{tot}(r) = \rho_{N}(r) + \rho_{H}(r)
\\&&\nonumber
\\&& \,\,\,\,\,\,\,\,\,\,\,\,\,\,\,\,= \frac{\rho_{0,N}\,r_{s,N}}{r[1+(r/r_{s,N})]^{2}} + \frac{M_{H}\,r_{s,H}}{2\pi\,r(r_{s,H} + r)^{3}}
\label{equ:totprofile} 
\end{eqnarray}

We have the option in SUPERBOX to either define the characteristic density for the NFW halo, $\rho_{0,N}$, or provide the total mass, $M_{N}$, in which case the code will calculate the density in equation (\ref{equ:totprofile}) by means of the following,

\begin{equation}
\rho_{0,N} = \frac{M_{N}}{4 \pi  r_{s,N}^{3}} \bigg(\ln(r_{s,N}+r) - \ln(r_{s,N}) - \frac{r}{r_{s,N}+r}\bigg)
\label{equ:nfwdensity} 
\end{equation}

In order to fully define the objects used in the simulations we must provide values for the following parameters;  NFW total mass ($M_{N}$) and scale length ($r_{s,N}$), the Hernquist total mass ($M_{H}$) and scale length ($r_{s,H}$), and finally the cut-off radii for both, $r_{c,N}$,  $r_{c,H}$.  These last two values are used by the simulation during setup to define some physical limit to the objects, since both potentials stretch to infinity.  In general, the virial radius of the object is used as an approximate guide for the cut-off radius.  In addition, it is necessary to define the number of particles that each component (Halo + Luminous) is modelled with.  In all our investigations, both the NFW and Hernquist components were modelled using $10^{6}$ particles each.  We model the central galaxy as a fixed analytic potential (which is in effect equivalent to observing the interaction in the rest frame of the central galaxy) and observe the effect that the interactions have on the satellite.  Although the use of a static potential eliminates the effect of dynamical friction, in this analysis we are concerned with the effect of rapid `fly-by' interactions and not long term evolution.  In addition, dynamical friction is more of a factor for many-orbit interactions, and we restrict ourselves to simulations lasting only a few interactions (i.e. a few pericenters), which is often less than two complete elliptical orbits, e.g. Fig.\,\ref{plot2} (left).  Moreover, replacing the static potential with a particle based object confirmed that the orbits are changed by dynamical friction by $\la$ 5 \% in these types of interaction.  The parameters used for the analytic potential are the same as for the particle-object and as such are simulated with the same expressions, (\ref{equ:totprofile}) and (\ref{equ:nfwdensity}), above.

The grid sizes discussed in \S2.1 are chosen to compliment the dimensions of the object.  For the inner grid it is necessary that the dimensions cover at least a couple of scale radii. The middle grid is chosen to encompass the object, and is set as the virial radius of the object plus some additional margin.  The outer grid of the simulation is always chosen to allow the objects full range of dynamical motion.  For this investigation the outer grid, and hence the simulation area, was set as 1Mpc $\times$ 1Mpc.

Models are set up in virial equilibrium making use of the Jeans formula as follows,

\begin{equation}
\langle v_{r}^{2} \rangle_{i} = \frac{1}{\rho_{i}(r)} \int_{r}^{r_{cut}} \frac{d \phi_{total}(r')}{dr'}\rho_{i}(r') dr'
\label{equ:jeans} 
\end{equation}

\noindent where  $\phi_{total}(r')$ is the combined potential of the NFW and Hernquist components.  Having $\langle v_{r}^{2}\rangle$ we generate velocities with a Gaussian distribution.

The objects are constructed using the procedure described in \citet{krou} and \citet{fellb}.  Once the grid sizes are chosen, and the parameters of the object are set, we allow the object to be integrated in isolation until relaxation with the grid sizes has occurred and the object has reached equilibrium.  This is necessary for all numerical realisations of stationary self-gravitating systems because discreteness of the model always implies an initial mismatch between kinetic and potential energy.  Equilibrium is determined by observing the fluctuation of the Lagrangian-radii of the mass shells of 10\% to 90\% of particles present in the simulation.  An example of this is shown in Fig.\,\ref{plot1} (right).   A stable flat line shows that no energy is entering or being bled from the system, and that the object has reached a stable equilibrium.  Once the objects have reached this point, they can then be used in the desired simulation scenario.  For example, Fig.\,\ref{plot1} (left) displays a typical contour density plot for a stable NFW that has undergone relaxation and is ready to be used in a simulation.  Also shown in Fig.\,\ref{plot2} (right) is an example of the same contour plot after undergoing an interaction scenario such as will be detailed in \S3.  We assume the interactions to be spherically symmetric and hence model the interactions in the x-y plane of the simulated area, and no prescription of adiabatic contraction is included when setting up the objects. 

%however we suspect that the types of changes that we are trying to observe should be greater in magnitude and operate on shorter time scales than those brought about through adiabatic contraction.

\begin{figure*}
\epsfig{file= 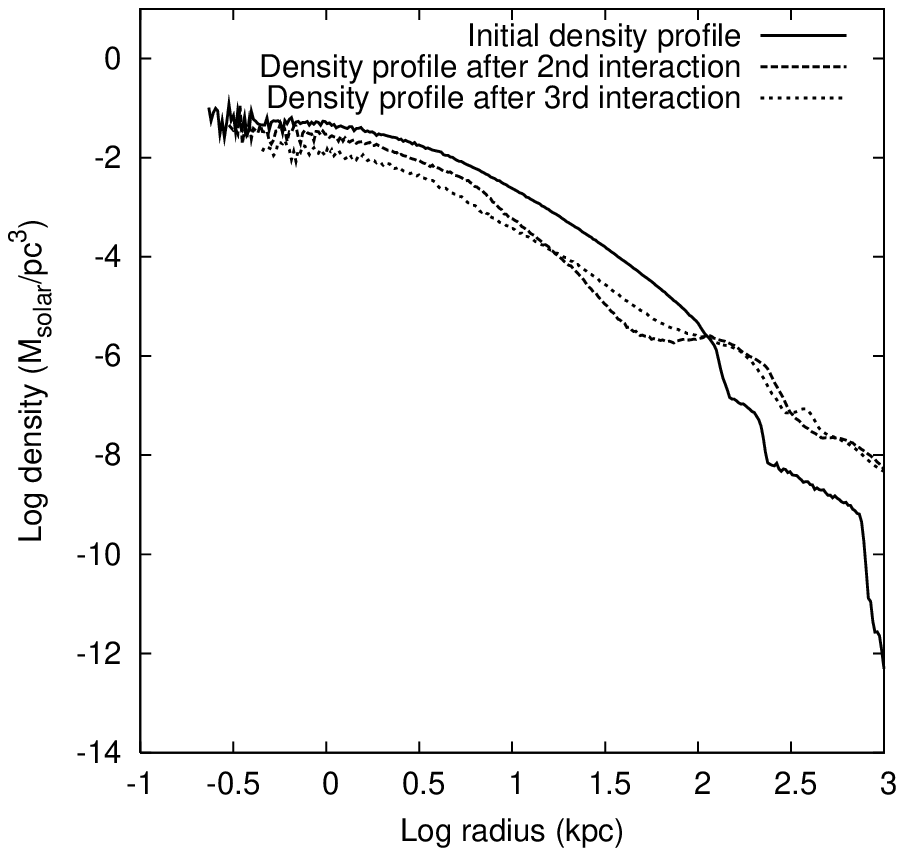,scale=0.64}
\epsfig{file=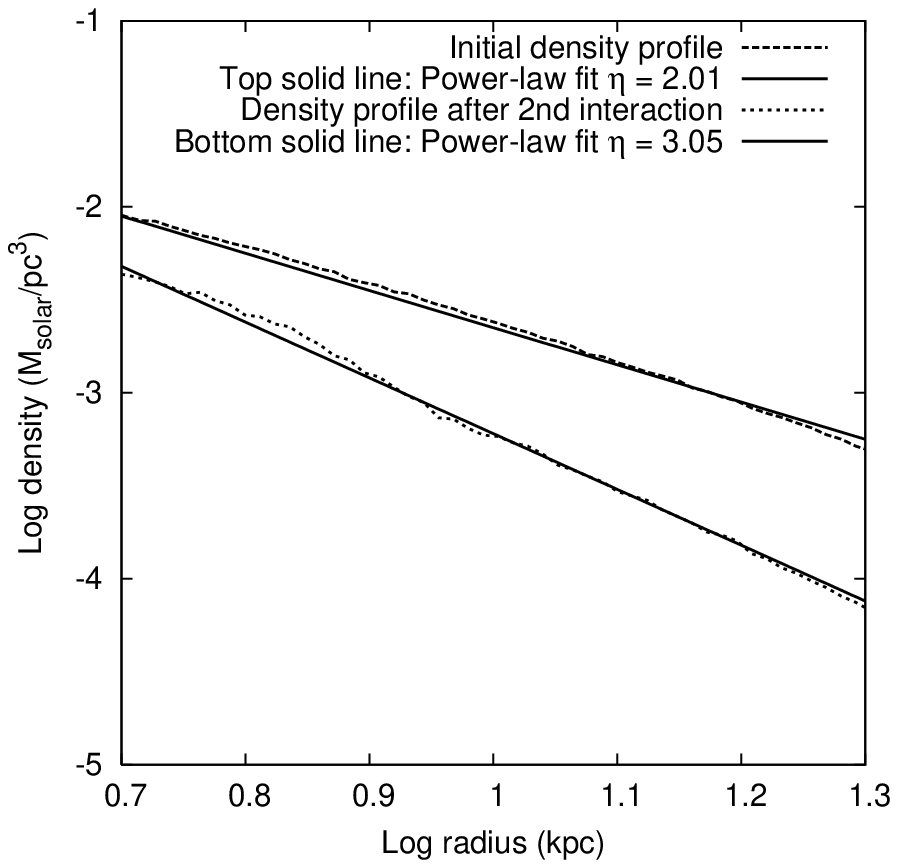,scale=0.64}
\epsfig{file=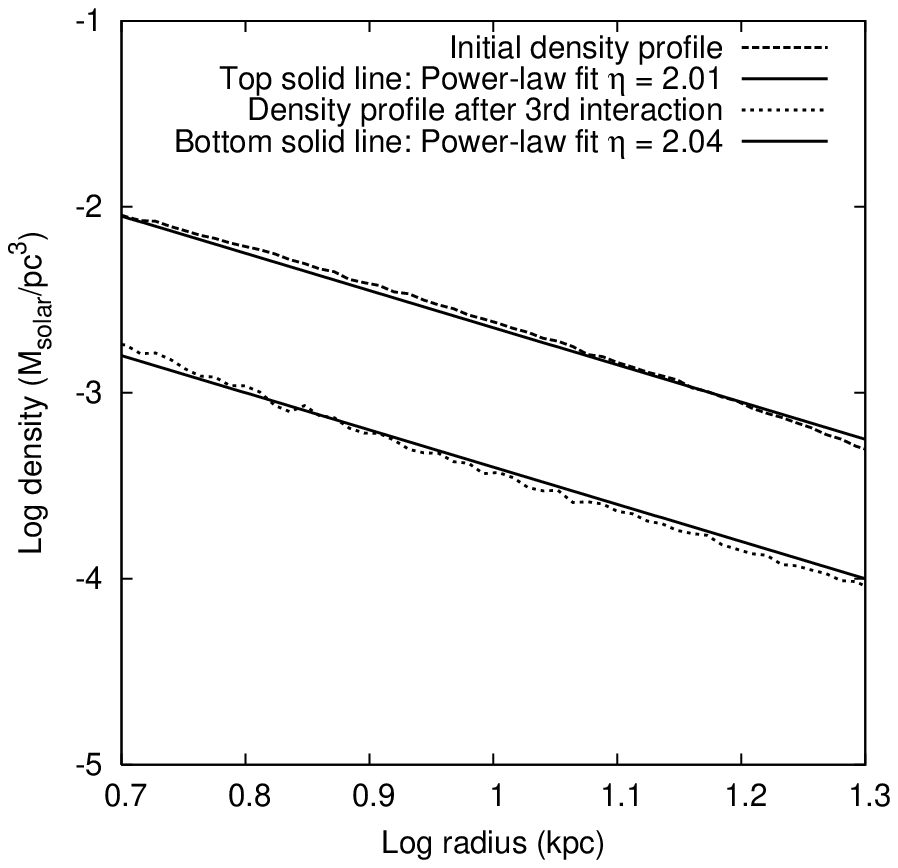,scale=0.64}
\caption{
{\sl Left:}  Combined density profile (NFW + Hernquist) before and after a 2nd and 3rd interaction between a $10^{11}$ $\rm M_{\sun}$ galaxy and $10^{12}$ $\rm M_{\sun}$ object modelled with an analytic potential.  Initial x-y position and velocity of the satellite were (0,-300) kpc and (+60,0) km\,s$^{-1}$ respectively. {\sl Middle:} Enlarged plot of the density profile in the region $\la$15kpc after the 2nd interaction.  Dashed line is the combined density profile before interaction, while dotted line is the combined density profile after interaction.   Top solid line shows a power-law fit with $\eta$ = 2.01 $\pm$ 0.03 i.e. isothermal, while bottom solid line shows a power-law fit with $\eta$ = 3.05 $\pm$ 0.07.  The profile has steepened significantly.  {\sl Right:} Enlarged plot of the density profile in the region $\la$15kpc after the 3rd interaction, $\sim$ 2 $\times$ $10^{9}$ years after the 2nd interaction.  Dashed line is the combined density profile before interaction, while dotted line is the combined density profile after interaction.  Top solid lines show a power-law fit with $\eta$ = 2.01 $\pm$ 0.03, with the bottom at $\eta$ = 2.04 $\pm$ 0.06.  The profile has returned to an isothermal state.}
\label{plot3}
\end{figure*}

\begin{figure*}
\epsfig{file=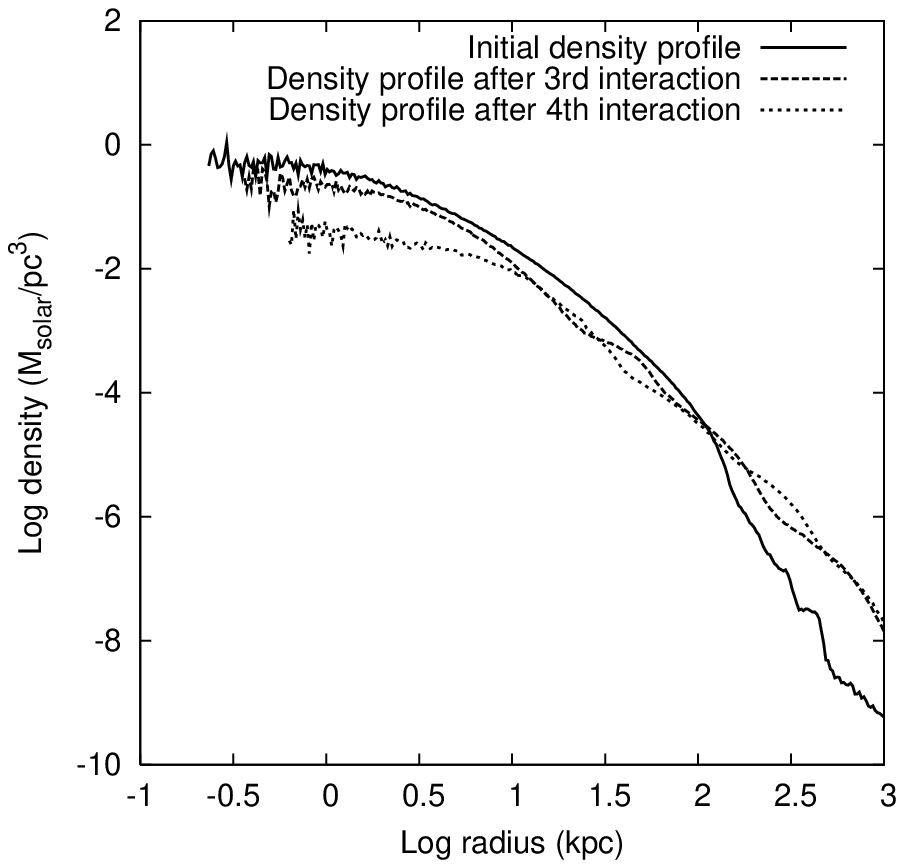,scale=0.64}
\epsfig{file=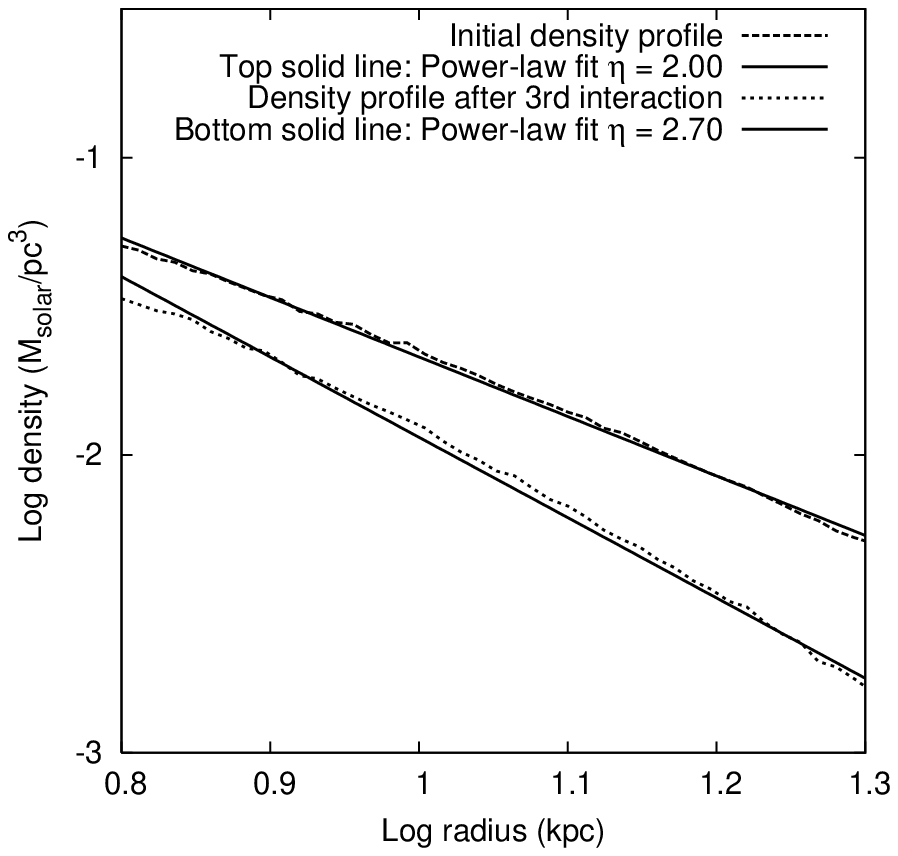,scale=0.64}
\epsfig{file=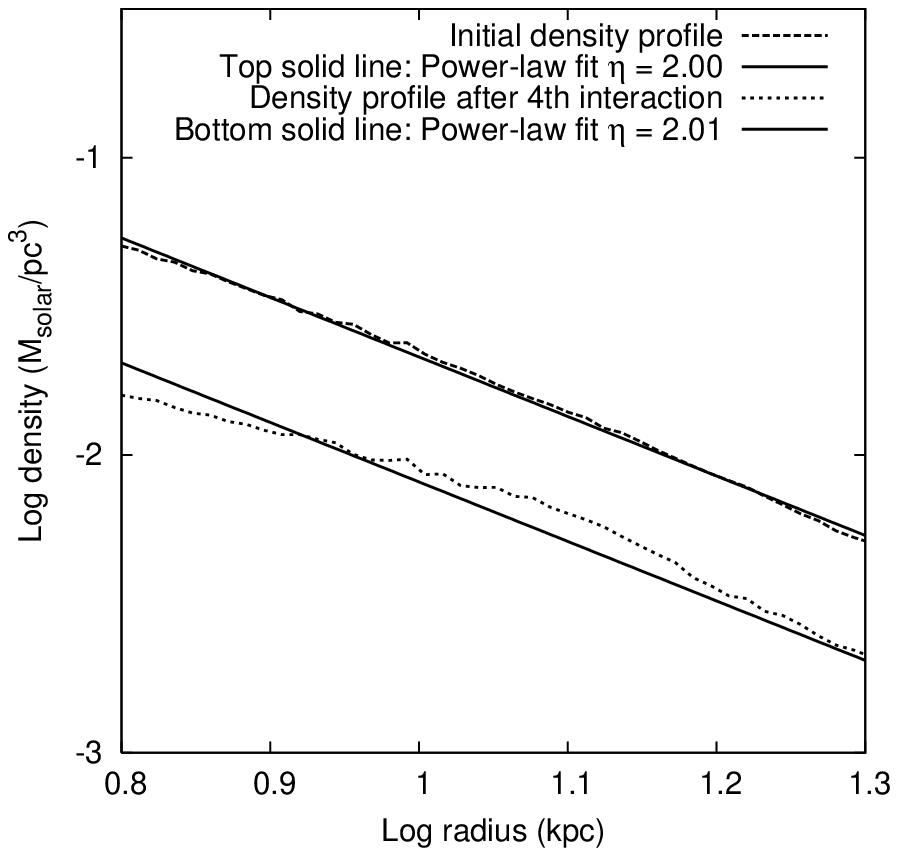,scale=0.64}
\caption{
{\sl Left:}  Combined density profile (NFW + Hernquist) before and after a 3rd and 4th interaction between a $10^{12}$ $\rm M_{\sun}$ galaxy and $10^{12}$ $\rm M_{\sun}$ object modelled with an analytic potential.  Initial x-y position and velocity of the satellite were (0,-300) kpc and (+100,0) km\,s$^{-1}$ respectively. {\sl Middle:} Enlarged plot of the density profile in the region $\la$15kpc after the 3rd interaction.  Dashed line is the combined density profile before interaction, while dotted line is the combined density profile after interaction.   Top solid line shows a power-law fit with $\eta$ = 2.00 $\pm$ 0.02  i.e. isothermal, while bottom solid line shows a power-law fit with $\eta$ = 2.70 $\pm$ 0.06.  The profile has steepened less so than in the case shown in Fig.\,\ref{plot3}, although the effect is still significant.  {\sl Right:} Enlarged plot of the density profile in the region $\la$15kpc after the 4th interaction, $\sim$ 0.5 $\times$ $10^{9}$ years after the 3rd interaction.  Dashed line is the combined density profile before interaction, while dotted line is the combined density profile after interaction.   Top solid lines show a power-law fit with $\eta$ = 2.00 $\pm$ 0.02, with the bottom at $\eta$ = 2.01 $\pm$ 0.20.  The profile has returned to an approximately isothermal state.}
\label{plot4}
\end{figure*}

\section{Analysis and Results}

We have performed a number of simulated interactions covering a variety of galaxy masses and initial orbits.  The mass range investigated is $\sim$ $10^{11}$ to $10^{13}$ $\rm M_{\sun}$.  In all interactions the central galaxy was located at (0,0) kpc in the x-y plane.  We vary the orbit of the satellite galaxy by changing both the initial starting position and tangential velocity (within the bounds of typical velocity dispersion for group members).  For the determination of our results we analyse the satellite features after modelling through a significant fraction of the Hubble time.  In most cases this involves several passes of the central galaxy.  An example of a typical orbital interaction path is displayed in Fig\,\ref{plot2} (left).  The central galaxy lies at (0,0) kpc, and the satellite galaxy begins at (0,-300) kpc with a tangential velocity of +100 km\,s$^{-1}$ in the x-dimension.  We also show 100 kpc and 150 kpc radial distances from the central potential, beyond which the satellite spends 78\% and 64\% of its orbit time respectively.  This provides some quantitative estimate and understanding for the probability of finding such a satellite galaxy away from the central galaxy and group centre.

We highlight results from two particular simulations.  The findings are non-unique to the particular set up and the results were found to be quite general for a variety of initial conditions.   The main results are displayed in Fig.\,\ref{plot3} and Fig.\,\ref{plot4}.  We display the density profile of the satellite before and after undergoing the interaction, then focus on the region $\la$15kpc as this dominates strong lensing. We fit the combined density profiles with a simple power-law model of the form,

\begin{eqnarray}
\rho_{tot}(r) = \rho_{0} \, r^{-\eta}
\label{equ:density} 
\end{eqnarray}

\noindent where the free parameters are $\rho_{0}$ (some normalisation constant) and $\eta$, the steepness of the profile.  For the fitting procedure itself we use equal sized logarithmic bins and employ a nonlinear least-squares (NLLS) Marquardt-Levenberg algorithm.

In Fig.\,\ref{plot3} (left) we show the initial combined density profile (NFW + Hernquist) for the simulated satellite galaxy (topmost line), and the density profile after the 2nd and 3rd interactions immediately below.  The masses of the interacting bodies were $10^{11}$ and  $10^{12}$ $\rm M_{\sun}$ for the satellite and central galaxy respectively. The initial mass-to-light ratio of both galaxies was set to 10:1, while the starting position was set at (0,-300) kpc, with an initial tangential velocity of +60 km\,s$^{-1}$ in the x-dimension.  For the purposes of this study we are interested in how the interactions in the outer part of the halo effect both the luminous and dark matter components in the inner part of the galaxy, particularly the region $\la$15kpc as this scale encompasses the strong lensing regime.  In Fig.\,\ref{plot3} (middle), we enlarge this region, just comparing the initial and post-2nd interaction density profiles.  Here, the dashed line is the profile before interaction, while the dotted line is the profile after interaction. The top solid line shows a power-law fit with $\eta$ = 2.01 $\pm$ 0.03, i.e. isothermal, while the bottom solid line shows a fit with $\eta$ = 3.05 $\pm$ 0.07.  We clearly see significant steepening occurring after the 2nd interaction over the region of $\sim$ 5 - 20 kpc.  The right plot of Fig.\,\ref{plot3} focuses on comparing the initial and post-3rd interaction density profiles.  The time elapsed between the two consecutive interactions is $\sim$ 2 $\times$ $10^{9}$ years, during which the slope of the profile remains approximately constant at $\eta$ $\sim$ 3.0.  We now see that after the 3rd interaction, the slope has returned once more to an isothermal state at $\eta$ = 2.04 $\pm$ 0.06, and hence the steepening observed after the 2nd interaction appears transient in nature.

\begin{figure*}
\centering
\epsfig{file=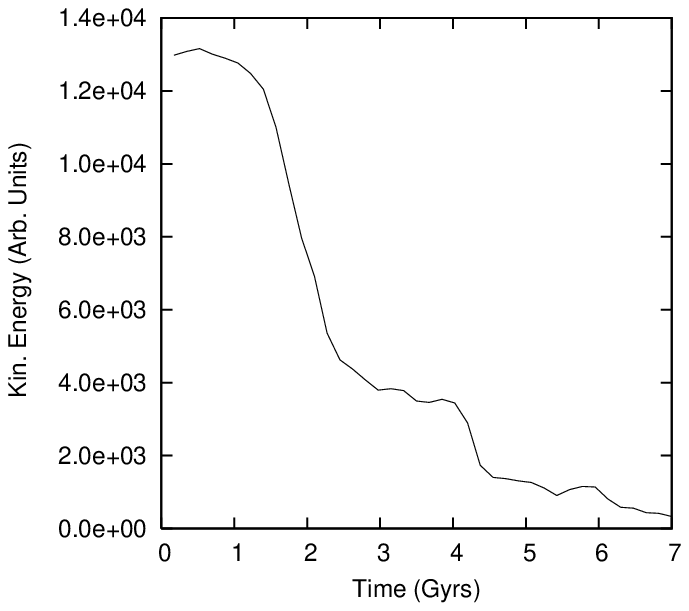,scale= 0.75}
\epsfig{file=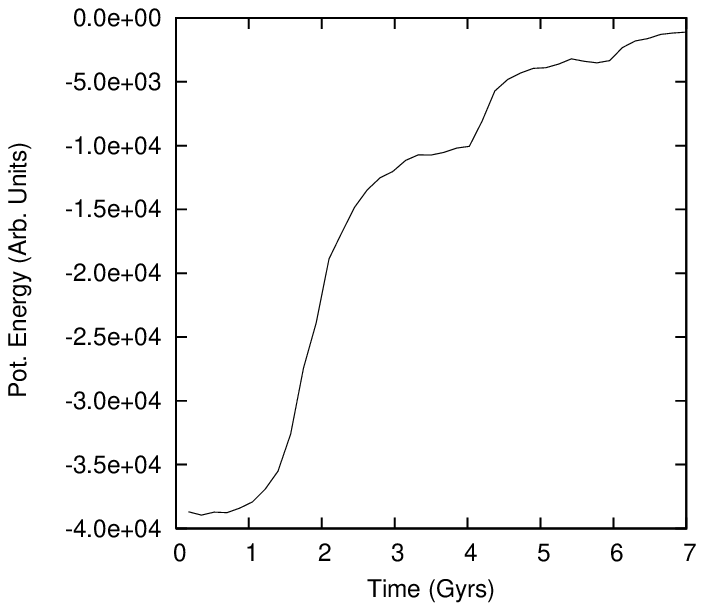,scale= 0.75}
\epsfig{file=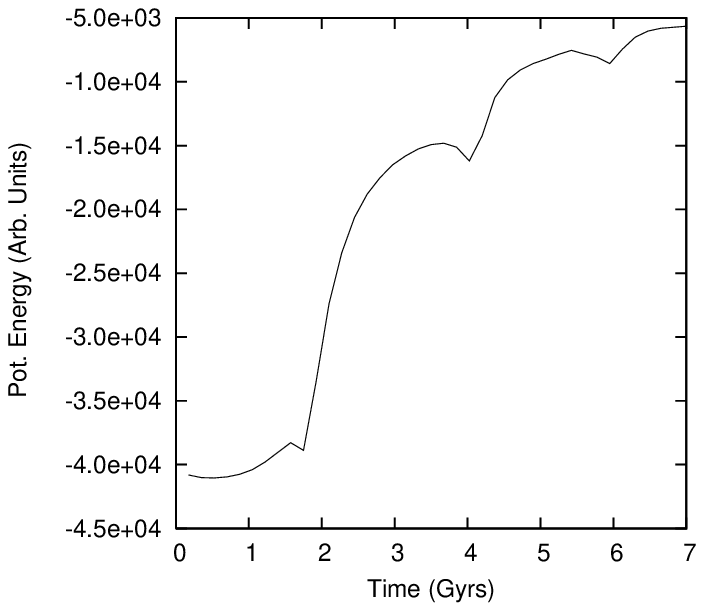,scale=0.75}
\caption{
{\sl Left:}  The time evolution of kinetic energy of the bound remnants of the interacting satellite of Fig.\,\ref{plot3}.  Note the three fluctuations in energy resultant from the three interactions and that this energy change becomes progressively smaller with time and hence the steepening of the profiles becomes less severe after each passage.  {\sl Middle:} The time evolution of potential energy of the bound remnants of the interacting satellite of Fig.\,\ref{plot3}.  Again we see the three energy fluctuations, each becomes less severe after each passage.  This results in the profile being ultimately unchanged despite a transient steepening during the first interactions. {\sl Right:} The time evolution of potential energy just within the region of 5-20kpc, for the case of the interacting satellite of Fig.\,\ref{plot3}.  We see the potential well increase in depth after the interactions, confirming the steepening of the profile, which again lessens after each interaction.}
\label{plot5}
\end{figure*}

We see a similar effect taking place in Fig.\,\ref{plot4}, except in this case we are investigating the interactions between two galaxies of similar mass, both being $10^{12}$ $\rm M_{\sun}$.  We see that the reduction in the relative mass necessitates a greater number of interaction events to initiate the steepening when compared with the previous case.  In the middle plot of Fig.\,\ref{plot4} the steepened profile after the 3rd interaction can now be fitted with a slope of $\eta$ = 2.70 $\pm$ 0.06 over the range of $\sim$ 6 - 20 kpc.  Although less than in the $10^{11}$ vs. $10^{12}$ $\rm M_{\sun}$ case, this is still a significant effect.  Once again, after the next interaction we observe the profile returning to an approximately isothermal state with $\eta$ = 2.01 $\pm$ 0.20 (Fig.\,\ref{plot4}, right).  This time the transient steepening lasts for $\sim$ 0.5 $\times$ $10^{9}$ years.

We now consider the steepening effect seen in Fig.\,\ref{plot3} and Fig.\,\ref{plot4} in terms of energy transfer during the interactions.  Displayed in Fig.\,\ref{plot5} is the time evolution of the kinetic and potential energies of the bound remnants for the interactions shown in Fig.\,\ref{plot3}.  During the fly-bys of the satellite galaxy, energy gets transformed from orbital to internal energy.  Particles in the satellite gain kinetic energy and those that are loosely bound, mainly in the outer parts, get stripped resulting in the steepening of the outer profile.  Closer towards the inner portion of the galaxy tidal shocks are more important than stripping (e.g. \citealt{read}; \citealt{gned}).  Close pericentre passes can generate matter compressions possibly resulting in the observed transient fluctuations in the density profile.  As the mass gets redistributed and the object returns once again to equilibrium, the original slope returns to its approximate initial state.  Notice in Fig.\,\ref{plot5} that this energy change becomes progressively smaller with time and hence the steepening of the profiles becomes less and less severe after each passage, resulting in a return to isothermality.  With greater time we see no change in the profile, confirming the transient nature of the steepening seen in Fig.\,\ref{plot3} and Fig.\,\ref{plot4}.  Ultimately, the density profiles will be approximately unchanged from their initial state, in agreement with a number of other studies (e.g. \citealt{kaza}).  In the righthand plot of Fig.\,\ref{plot5}, we display the evolution of potential energy with time for just the 5-20 kpc region.  Notice how the potential well increases in depth after the interactions, confirming the steepening of the profile that is seen in Fig.\,\ref{plot3}.

\begin{figure}
\centering
\epsfig{file=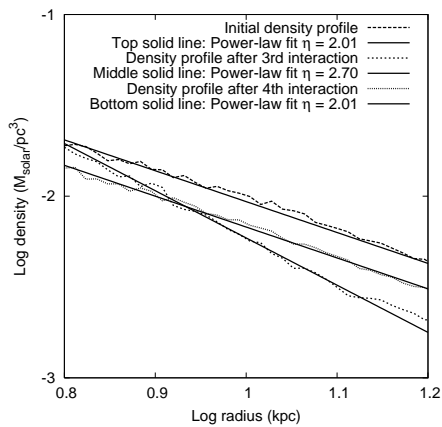,scale=0.64}
\caption{Density profile before and after a 3rd and 4th interaction between two $10^{12}$ $\rm M_{\sun}$ objects.  Here we have replaced the static potential (as in Fig.\,\ref{plot3} and Fig. \,\ref{plot4}) with a particle based object.  As in the static potential case, we continue to see the distinctive steepening followed by a return to an approximate isothermal state.}
\label{plot6}
\end{figure}

For both of the interaction scenarios displayed in Fig.\,\ref{plot3} and Fig.\,\ref{plot4}, we also investigated the effect of varying the initial mass-to-light ratio of the satellite galaxy.  In addition to the cases shown for an M/L = 10:1, we performed simulations using M/L = 20:1 and 5:1.  In all cases there was very little variation in the results, and the overall effect of steepening followed by a return to isothermality was observed.  We also examined higher mass interactions between $10^{12}$ and $10^{13}$ $\rm M_{\sun}$ objects and again observed a similar pattern of steepening followed by a return to isothermality, although the larger of these two objects represents a mass normally only associated with central locations in clusters rather than in groups.  To confirm that the exclusion of dynamical friction does not overly effect the fundamental result, we replace the central static potential with a particle modelled object.  We perform  a simulation between two $10^{12}$ $\rm M_{\sun}$ galaxies, i.e. the case in which dynamical friction should be at its greatest.  As shown in Fig.\,\ref{plot6} we continue to see the transient steepening in the density profile.

\section{Discussion and Conclusions}

We appear to require steeper than isothermal density profiles in order to reconcile certain lens galaxies with current estimates of the Hubble constant from HST/WMAP.  These galaxies include PG1115+080, B1600+434, HE2149-2745, and SBS1520+530.  One possible explanation is that certain early-type lens galaxies are offset from the Fundamental Plane and as such early-type galaxies in general might display a heterogeneity in their structure and hence mass density profiles.  This theory proposes that the heterogeneity results from group interactions between the lens galaxy and its fellow group members, resulting in central galaxies having isothermal profiles and satellites having steeper than isothermal profiles.  A number of studies seem to agree that the discrepancies in derived $H_{0}$ values could be the result of structural differences in lens galaxies as opposed to other factors \citep{treub,kochb}.

We have investigated group interactions between group members with masses ranging from $10^{11}$ to $10^{13}$ $\rm M_{\sun}$.  Our investigations show a significant steepening of the density profile in the inner region, pertinent to strong lensing.  This effect appears to be independent of the initial mass-to-light ratio, while the precise initial orbital parameters made little difference to the fundamental result provided consecutive interactions with the central galaxy took place.  The magnitude of the steepening itself varied with the relative masses of the interacting objects;  for cases where the mass differed by an order of magnitude, as for the cases in interactions involving $10^{11}$ vs. $10^{12}$ $\rm M_{\sun}$ and $10^{12}$ vs. $10^{13}$ $\rm M_{\sun}$, we observe originally isothermal profiles ($\eta$ = 2) steepen up to $\eta$ = 3.  For the cases where masses were of similar magnitude, i.e. an interaction between two $10^{12}$ $\rm M_{\sun}$ group members, we observe a steepening of the profile slope up to $\sim$ $\eta$ = 2.7.  In both cases, modelling such systems with isothermal models would result in a significant underestimate of the Hubble constant.

We stressed in \S1 that most early-type galaxies appear to reside in groups \citep{tonry, fass, will}, and as discussed by \citet{kochb}, there is a play-off in lensing optical depth between satellite galaxies being more numerous than the most massive central group galaxies, yet having a lower lensing cross-section (see also \citealt{ogur}).  Even for satellite galaxies, the steepening in the inner region seems transient in nature, with consecutive interactions often returning the profile to an isothermal state within a timeframe of $\la$ 2 Gyr.  This factor may help explain why lens galaxies that produce lower values of $H_{0}$ (i.e. those with possibly steeper profiles) are far fewer in number than those which agree with both the HST key project value for $H_{0}$ and isothermality.  In addition, it is important to note that this transient fluctuation in steepness is consistent with studies showing that density profiles over longer time scales can be robust despite strong interactions or mergers \citep{kaza, kaza2}.

We should highlight that other factors besides the environment discussed in \S1 are thought to contribute to discrepancies in derived Hubble constant values from certain systems.  Recently, \citet{sahab} showed in an investigation with models of 35 galaxy lenses that shape-modelling degeneracies (e.g. caused by triaxiality) can also contribute to changes in the time delays and hence the derived Hubble constant value.  This effect alone would need to be very significant to fully account for the systems with the lowest derived values of $H_{0}$.  It seems probable that a combination of effects, along with neglect of physical steepening due to stripping, can account for low $H_{0}$ values.  

Looking towards the future, space and ground based observatories such as GAIA and LSST will uncover roughly fifty times more multiply imaged quasars than we have identified today.  With a greater sample of lens systems to study, and by combining lens models with stellar dynamical constraints,  heterogeneity in early-type galaxies as a function of environment and redshift will be explored in detail.
This will also allow unprecedented constraints on the assembly and distribution of dark matter on galaxy and galaxy group scales. 

\section*{Acknowledgments}
We thank Mark Wilkinson and Chris Fassnacht for discussions and the anonymous referee for their constructive comments.  This work was supported by PPARC through a PhD studentship (BMD), by the Royal Society (LJK), and by PPARC (MF)

\end{document}